\documentclass{elsart}

\usepackage{graphics}
\newcommand{\Closure}{\mbox{${C}$}}

\newcommand{\Yield}{\mbox{$\Upsilon$}}

\newcommand{\gammavec}{\mbox{${\boldmath \gamma}$}}

\newcommand{\szz}{\mbox{$\sigma_{\mathrm{zz}}$}}
\newcommand{\sxx}{\mbox{$\sigma_{\mathrm{xx}}$}}

\newcommand{\sxz}{\mbox{$\sigma_{\mathrm{xz}}$}}

\newcommand{\sign}{\mbox{sign}}

\begin{document}
\begin{frontmatter}
\title{Stress Propagation in Sand}
\thanks{}

\author{M.~E.~Cates and J.~P.~Wittmer}
\address{
Department of Physics and Astronomy, University of Edinburgh\\
King's Buildings, Mayfield Road, Edinburgh EH9 3JZ, UK
}

\date{1 April  1997}
\setcounter{page}{1}
\begin{center}
PACS numbers 46.10.+z, 46.30.-i, 81.35.+k, 83.70 Fn
\end{center}

%%%%%%%%%%%%%%%%%%%%%%%%%%%%%%%%%%%%%%%%%%%%%%%%%%%%%%%%%%%%%%
\begin{abstract}
We describe a new continuum approach
to the modelling of stress propagation in static granular media,
focussing on the conical sandpile created from a point source.
We argue that the stress continuity equations should be closed
by means of scale-free, local constitutive relations
between different components of the stress tensor,
encoding the construction history of the pile:
this history determines the organization of the grains,
and thereby the local relationship between stresses.
Our preferred model FPA (Fixed Principle Axes) assumes that
the eigendirections (but not the eigenvalues) of the stress tensor
are determined forever when a material element is first buried.
Stresses propagate along a nested set of archlike structures within
the medium; the results are in good quantitative agreement with
published experimental data.
The FPA model is one of a larger class, called OSL
(Oriented Stress Linearity) models, in which the direction of
the characteristics
for stress propagation are fixed at burial.
We speculate on the connection between
these characteristics and the {\em stress paths} observed microscopically.
\end{abstract}
\end{frontmatter}
%%%%%%%%%%%%%%%%%%%%%%%%%%%%%%%%%%%%%%%%%%%%%%%%%%%%%%%%%%%%%%

\section{Introduction}\label{sec:Intro}
A sandpile is normally constructed by pouring sand from a stationary
point source. Each element of
sand arrives at the apex of the pile, rolls down the slopes, comes to rest,
and is finally buried. Thus successive layers at the angle of repose
$\phi$ are added to the symmetrical cone as the height $H$ of the pile
increases. Some of the simplest questions one can ask about
this system concern the distribution of stresses in the pile.
Intuitively one would guess that the maximum vertical force
would be recorded directly beneath the apex of the pile;
but in fact, the experiments show a pronounced {\em dip\/} in the force
beneath the apex.\cite{smidnovosad,otherdip}

\begin{figure}
\centerline{\scalebox{.4}{\includegraphics{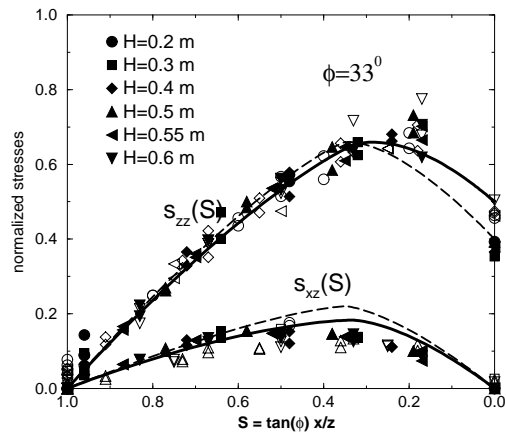}}}
\caption[]{
Rescaled data of Smid and Novosad\cite{smidnovosad} from conical piles
of different heights ($H=20-60 cm$) for quartz sand (closed symbols)
and NPK-1 fertilizer (open symbols).
%These cohesionless Coulomb materials materials have different densities
%but roughly the same repose angles $\phi \approx 33\pm 1^o$.
Radial stress field (RSF) scaling is obeyed to experimental
accuracy.
The data are compared
with two versions of the FPA model in three dimensions:
the bold line assumes an {\em uniaxial\/} stress tensor,
the dashed one a possible alternative discussed in the text.
%there are no adjustable parameters in the FPA model.
\label{fig:smidnovosad}}
\end{figure}

In fig.~\ref{fig:smidnovosad} we show vertical normal and shear stresses
measured by Smid and Novosad \cite{smidnovosad} on the supporting surface
for piles of varying height for two different cohesionless Coulomb materials.
We normalized the stresses by the total weight of the piles;
notice the good ``radial stress field'' (RSF) scaling of curves
from piles of different heights. These data show that to experimental accuracy,
there is {\em no intrinsic length scale\/} in the sandpile problem!
This first observation is one of the cornerstones in our modelling approach
to the stress distribution in aggregates consisting
of cohesionless hard particles held up by static frictional forces.
\cite{nedderman,coulomb}
The stresses were measured on a linear array of sensors across the width
of the piles; there are two data points for each symbol corresponding
to two opposite points on the pile. Their difference gives an indication
of the size of the fluctuations: Stochastic processes occuring in the
propagation of stresses are apparently not so broad as to prevent
one formulating the ``continuum limit'' for piles of reasonable size.
This second observation strongly suggests a traditional continuum approach,
\cite{sand1,sand2}, although there is certainly some noise.
\cite{dantu,Liu,coppersmith,sand7}

\section{Systematic Continuum Approach}\label{sec:Continuum}

%\subsection{Hydrodynamic limit}
{\em Continuum limit:}
We suppose that after sufficient coarse-graining
the force propagation can be described by a
traditional ``continuum limit'' based on the continuity
equation of the stress tensor:
\begin{equation}
\nabla_i\sigma_{ij} = g_j
\end{equation}
where the source term $g_j$ is the gravitational field (we take
units where the density is unity).
This equation is incomplete:
in $d$ dimensions $d (d-1)/2$ {\em constitutive relations\/} (CR)
are required. These should be based on physically motivated assumptions.
For simplicity we will mostly stick to the $d=2$ dimensional case
which requires one CR.

\begin{figure}
\centerline{\scalebox{0.4}{\includegraphics{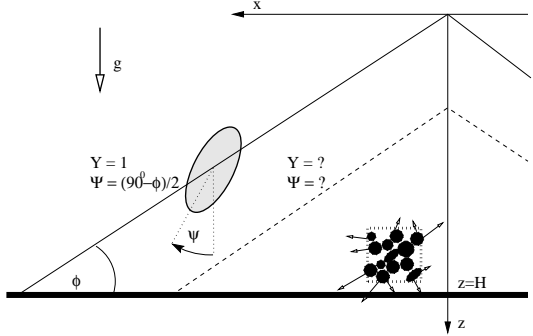}}}
\caption[]{A two dimensional symmetrical PS-pile of Height $H$.
The $(x,z)$ coordinates used are shown:
The $z$-axis points from the apex down in the
direction of gravitation.
The cohesionless Coulomb material is assumed to be at IFS
fixing two invariant quantities near the surface:
the ellipticity of the stress tensor $\Yield=1$ and its direction $\Psi$.
The major principal axis then bisects the surface and the vertical.
Nothing is known {\it a priori} however on both quantities
on former surfaces (broken line). The invariant quantity
memorized by the texture will provide the missing CR.
\label{fig:boundary}}
\end{figure}

%\subsection{Coulomb boundary conditions}
{\em Coulomb boundary conditions:}
Let us first fix the boundary
conditions on free surfaces.
Because of the symmetry of the piles we can restrict our attention to
the left side (fig.~\ref{fig:boundary}).
The Coulomb yield criterion (that shear forces do not exceed the maximum
permitted by static friction) can be written in terms of the normalized
ellipticity function
\begin{equation}
\Yield \stackrel{\mathrm{def}}{=}
\frac{1}{\sin(\phi)} \frac{\sigma_1-\sigma_2}{\sigma_1+\sigma_2}
\label{eq:Yield}
\end{equation}
where $\phi$ is a material parameter and $\sigma_1,\sigma_2$ are major and
minor principal stresses. In fact, it is easily established \cite{coulomb}
that $\phi$ is the angle of repose for a freestanding pile whose surface
is a slip plane (such as a pile created from
a point source). This IFS (incipient failure at surface) boundary condition
requires (a) that the yield function $\Yield = 1$ on the surface; and (b)
that the direction of the major principal axes bisects the vertical and the
surface:
$\Psi = (90^0-\phi)/2$ where the notation is shown in fig.~\ref{fig:boundary}.
Note that the Coulomb criterion does not (without further assumptions)
fix the stresses in the bulk; here we know only
that $\Yield \leq 1$.

%\subsection{No intrinsic length}\label{sec:NointrinsicLength}
{\em No intrinsic length:}
In fig.~\ref{fig:smidnovosad} we have shown strong evidence
that the stresses scale with height.
We assume therefore that there exists no intrinsic length scale.
This severely limits the possible form the missing constitutive relation
can take.
In particular it rules out all elastic models invoking a relation
between stress and strain: associated with a stress scale (Young modulus)
there is always (under gravity) a length scale.
Effectively we require that this length scale
is much larger than the size $H$ of the pile and can be sent to infinity.
Quantities like the direction of the tensor and its ellipticity have to
scale like $\Psi(x,z)=\Psi(S)$, $\Yield(x,z)=\Yield(S)$
where $S=\tan(\phi) x/z$ is a scaling variable
(defined to be unity on the free surface). This is called RSF (radial
stress field) scaling.

%\subsection{Locality}\label{sec:Locality}
{\em Locality:}
Now we seek a CR,  which
(i) is scale-free (no intrinsic length or stress scale),
(ii) may depend on the local packing of the grains fixed by the
aggregation history,
and (iii) is simple. We therefore assume the {\em locality\/} of the CR:
the {\em relation between stresses}
is unaffected by distant loads, although the stresses themselves do of
course depend on such loads.
For this to hold, we should check {\em a posteriori\/} that no yield occurs.
In two dimensions this assumption, together with the earlier
hypotheses requires that there is a function \Closure, such that
\begin{equation}
\Closure\left(\Yield,\Psi,S\right)=0.
\label{eq:Locality}
\end{equation}

%\subsection{Perfect memory}\label{sec:Memory}
{\em Perfect memory:}
For a pile made from a point source in two dimensions, it is intuitively
clear that
particles on the left of the apex may transmit stress anisotropically
due to packing features {\em memorized\/} from the
deposition,
with particles on the right the mirror image.
Particles exactly under the apex have never rolled down either
slope and a discontinuity of transmission properties can be expected there.
We take this history-dependence into account in the simplest way possible,
by arguing that \Closure\ for a volume element is ``frozen in''
at burial:  the {\em Perfect Memory} assumption.
Since each volume element is buried with the same value $S=1$ this rules
out any dependence of \Closure\ on the magnitude of $S$:
$\Closure\left(\Yield,\Psi,\sign(S)\right)=0$.

\begin{figure}
\centerline{\scalebox{.35}{\includegraphics{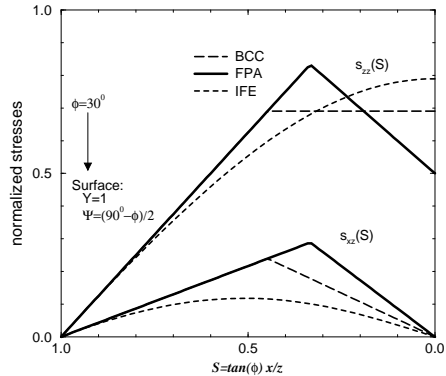}}}
\caption[]{Reduced shear stress and reduced normal stress
as a function of scaling variable $S=\tan(\phi) x/z$ for a
symmetrical two dimensional sandpile with a surface obeying IFS
with an angle of repose $\phi=30^0$.
Results for the IFE model (found numerically), the FPA model, and the
BCC model are compared.
\label{fig:sa30D2}}
\end{figure}

%\subsection{Choice of constitutive relation}
{\em Choice of constitutive relation:}
For a pile constructed from a point source, the main consequence
of the perfect memory assumption is that (in two dimensions)
there exists one invariant property of the stress tensor,
depending on the texture as created during deposition,
that remains constant after additional layers are added.
Traditionally it is (often) supposed that the material remains
everywhere at incipient failure ({IFE}),
i.e. the CR is that $\Yield=1$ everywhere.
Numerical solution of the differential equations \cite{sand2}
shows that this gives a stress hump
in the middle of the pile (fig.~\ref{fig:sa30D2}) rather than a dip.
The same holds in $d=3$.
An alternative CR, used recently by Bouchaud, Cates and Claudin (BCC)
\cite{BCC} is to assume $\sigma_{xx}/\sigma_{zz}$ is a constant. This
closure does not break the symmetry between the left
and right halves of the pile; the CR is smooth crossing the central axis.
BCC gives, in two dimensions, a flat plateau in the stress.

%\subsection{FPA model}
{\em FPA model:}
More successful is the FPA model. In this model we assume that it is
not $\Yield$,
but the orientation of the stress tensor $\Psi$ which remain fixed
from the time of burial. This gives a dip (fig.~\ref{fig:sa30D2}).
In three dimensions, for a pile with rotational symmetry, we need one
additional CR
to close the problem. There are some well-known candidates for this
secondary closure, including equality of the radial and tangential
normal stresses, or that the stress tensor is uniaxial ($\sigma_2 = \sigma_3$).
Fortunately, the results do not depend much on which secondary closure is
used; both choices are compared with the experimental data in
fig.~\ref{fig:smidnovosad}. There are no adjustable parameters;
the agreement, though not perfect, is remarkably good.

\section{From Principal Lines of Force to FPA and OSL}\label{sec:FPA}
In order to motivate the FPA assumption from a
more microscopic point of view,
let us first step back from our {\em tensorial\/} approach to
the {\em scalar\/} model due to Edwards and coworkers \cite{Edwards}
which itself is motivated by the ``stress paths'' seen in
experiment.\cite{dantu,Liu,coppersmith}
In this model the force arising from an additional
small load travels along a straight line of constant angle
toward the base of the pile; lines in the left and right half are mirror images.
These ``principal lines of force" make up
a series of nested arches along which the weight is propagated.
It can easily be shown\cite{sand2} that this model has several drawbacks
(mechanically unstable, overprediction of dip, etc.).
On the other hand it expresses intuitively the
sound physical concept of local rules which tend -- due to
the memorized history -- to propagate stresses outwards.

\begin{figure}
\centerline{\scalebox{.35}{\includegraphics{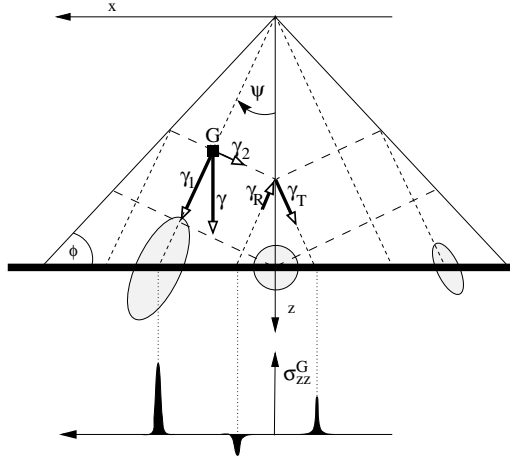}}}
\caption[]{Sketch of the geometry of the FPA model.
The stress ellipsoid has fixed inclination angle $\Psi = (90^0-\phi)/2$;
its ellipticity varies from zero at the centre of the pile
to a maximum in the outer region.
The outward and inward stress propagation
characteristics are indicated by short-dashed and long-dashed lines;
these are at rightangles and coincident with the principal axes of the stress
ellipsoid. The forces propagating from an additional source term
\gammavec\ at $G$ along the network of characteristics are shown
along with the resulting vertical forces $\sigma_{zz}^G$ on the bottom plate.
\label{fig:FPA}}
\end{figure}

Similar physics arises in our FPA model.
To see how this happens, we need to consider the characteristic rays
for stress propagation. These can be found by combining the stress
continuity equation
with the FPA closure
in cartesians, which reads
\begin{equation}
\frac{\sxx}{\szz} =  \eta  + \mu \frac{\sxz}{\szz}
\label{eq:OSL}
\end{equation}
with the coefficients $\eta=1$ and $\mu=-2\cot(2\Psi)$.
(Note that $\mu$ changes sign on the symmetry axis.)
The result is a wave equation in two dimensions, for which the
characteristics are straight
lines. In the FPA model, the directions of these
characteristics  are the same everywhere within each half of the pile, 
and {\em coincide with the principal axes themselves}.
This is shown in fig.~\ref{fig:FPA}. The discontinuity in the orientation
of the characteristics at the centre line leads to nontrivial reflection
and transmission rules there. \cite{sand2}

The physical outcome is that the stresses in the FPA model indeed propagate
down a set of nested arches in the sense that there are no shear forces
acting between one arch and its neighbours. However, there are normal forces,
without which the outer (incomplete) ``arches" would fall down.
This is an appealing mechanical picture. However, it still does not
explain why it is the principal axes that get remembered when an element
of sand is buried.

%\section{Fixed Characteristics (OSL)}\label{sec:OSL}
{\em Fixed Characteristics (OSL)}
At this stage we take a step back in our set of assumptions,
and generalize the FPA model slightly.
There the characteristics coincide with the principal axes and hence
are orthogonal.
In the broader class of OSL models (from ``Oriented Stress Linearity'')
orthogonality is not assumed but the constraint of fixed characteristics is
retained. In terms of Cartesian coordinates the OSL assumption
leads to linear relationships between reduced stress components,
as in eq.~(\ref{eq:OSL}), but with free coefficients $\eta$ and $\mu$.
But the repose angle can be found in terms of these \cite{sand2}, so the
OSL model has one free parameter (say $\eta$). 
This fixes the anisotropy on the
centre line of the pile (where $\sigma_{xz}=0$ by symmetry) and is unity for
FPA. (This allows the stress tensor to have simultaneously
the principal axes pertaining
to the right and left halves of the pile.)
In the general OSL case, the principal axes vary smoothly through
the centre-line, though the characteristics do not. Experimentally
fig.~\ref{fig:smidnovosad} shows that the data is well-fit by $\eta$
values close to unity. \cite{sand2}
A possible physical interpretation of OSL connects the
characteristics with the average directions of the stress paths.
In this view the {\em mean directions\/} of these stress paths
remain fixed at the time of burial and is unaffected by distant loads.
In the continuum limit these average directions become
characteristics of the stress propagation
equations, which OSL takes to be fixed.
This feature is not shared by, for example,
the IFE model and is not obvious. However, it is a testable hypothesis
(for example by simulation \cite{BWC1})
whose confirmation would go a long way to confirming the modelling
strategy we have used. It is important that we require only the
{\em average} directions of the stress paths to be fixed; it seems
likely that individual paths might be subject to strong noise effects
\cite{CB1,sand7}.

\section{Conclusion}\label{sec:Conclusion}
We have described a new continuum approach
to the modelling of stress propagation in static granular materials
composed of hard grains in frictional contact.
We reject traditional elasto-plastic
concepts in favour of a local, scale-free constitutive relation
among stresses which ``memorizes" invariant properties of the stress
tensor. These we associate with the local packing of
grains: {\em the constitutive equation encodes the construction
history}.
For a pile construced from a stationary point source, we favour the
FPA model where the eigendirections of the stress tensor ellipsoid
are fixed at the moment of burial. This surprisingly simple model gives
good quantitative agreement with published experimental results.
There is no adjustable parameter in this model. Although we have
no deep justification for the FPA assumption, it is one of a broader
class of models (OSL) for which the characteristics of stress
propagation are fixed at burial.  This suggests an appealing link
with the concept of microscopic stress paths; the OSL model would
suppose that the mean orientation of such paths is invariant under
loading.\cite{sand7}

\begin{ack}
%\section*{Acknowledgments}
The authors are indebted to their collaborators J.-P. Bouchaud and P. Claudin
who were closely involved in the work reviewed above. They also thank
S.F.~Edwards, R.M.~Nedderman and J.M.~Rotter
for illuminating discussions.
This work was funded in part by EPSRC under
Grant GR/K56223 and in part by the Colloid Technology programme.
\end{ack}

\end{document}